\documentclass[conference]{IEEEtran}
\IEEEoverridecommandlockouts
\usepackage{cite}
\usepackage{multirow}
\usepackage[pdftex]{graphicx}
\usepackage[cmex10]{amsmath}
\usepackage{amssymb}
\usepackage{caption}
\usepackage{bm}
\usepackage[inline]{enumitem}
\usepackage{gensymb}
\usepackage{tabu}
\usepackage{pifont}
\usepackage{csquotes}

\usepackage[pagebackref=true,breaklinks=true,letterpaper=true,colorlinks,bookmarks=false]{hyperref}

\newcommand{\etal}{\textit{et al}. }
\newcommand{\eg}{\emph{e.g}. }

\newcommand\blfootnote[1]{%
	\begingroup
	\renewcommand\thefootnote{}\footnote{#1}%
	\addtocounter{footnote}{-1}%
	\endgroup
}

\begin{document}

\title{\emph{Volenti non fit injuria}: Ransomware and its Victims}

\author{
\IEEEauthorblockN{Amir Atapour-Abarghouei,
Stephen Bonner and
Andrew Stephen McGough}

\IEEEauthorblockA{School of Computing, Newcastle University, Newcastle, UK  \\ \{amir.atapour-abarghouei, stephen.bonner3, stephen.mcgough\}@newcastle.ac.uk} 

}

\maketitle

\begin{abstract}

    With the recent growth in the number of malicious activities on the internet, cybersecurity research has seen a boost in the past few years. However, as certain variants of malware can provide highly lucrative opportunities for bad actors, significant resources are dedicated to innovations and improvements by vast criminal organisations. Among these forms of malware, ransomware has experienced a significant recent rise as it offers the perpetrators great financial incentive. Ransomware variants operate by removing system access from the user by either locking the system or encrypting some or all of the data, and subsequently demanding payment or \emph{ransom} in exchange for returning system access or providing a decryption key to the victim. Due to the ubiquity of sensitive data in many aspects of modern life, many victims of such attacks, be they an individual home user or operators of a business, are forced to pay the ransom to regain access to their data, which in many cases does not happen as renormalisation of system operations is never guaranteed. As the problem of ransomware does not seem to be subsiding, it is very important to investigate the underlying forces driving and facilitating such attacks in order to create preventative measures. As such, in this paper, we discuss and provide further insight into variants of ransomware and their victims in order to understand how and why they have been targeted and what can be done to prevent or mitigate the effects of such attacks.

\end{abstract}


\section{Introduction}
\label{sec:introduction}

With the growing influence of automated data handling systems in various aspects of the daily life of an average citizen, such as banking, education, health care and many other public and private services, systems security is becoming ever more important. Various malicious online activities by large criminal syndicates or independent individual bad actors now threaten any operation reliant on computer systems.

Of the numerous strains of malware regularly appearing online, ransomware is now of particular interest to the cybersecurity community \cite{zhang2019classification} as it is capable of targeting any user indiscriminately and can inflict irreversible harm on its victims. A ransomware often exploits low-level operating system mechanisms or security-based operations, such as cryptography, to isolate users from their assets (be these data services or systems), partially or on the whole. The user can only regain access if and when a sometimes-hefty \enquote{ransom} is paid, and in many instances, the access to the data is never returned to the user even if the ransom is paid in full \cite{moore2016detecting}. Consequently, due to the significant financial gain ransomware can offer the perpetrators \cite{laszka2017economics}, considerable resources are often put behind the creation of new and innovative variants, allowing them to bypass state-of-the-art anti-virus and anti-malware software \cite{kok2019ransomware}.

Initially, ransomware attacks seem to have been in the form of \emph{spray-and-prey}, with little targeting towards any particular individual. However, more recently, attackers have been moving towards more targeted attacks~\cite{fbi} -- the so-called \enquote{big game hunting}, in which cybercriminals target high-value organisations, putting significant value into identifying more targeted routes of entry.

As a result of this perceived high value for cybercriminals, the cybersecurity community has had to stay vigilant to maintain the ability to detect and avert constantly-emerging ransomware attacks. Malware activities have conventionally been identified either at the network level \cite{gu2008botminer, cabaj2018software}, system level \cite{bayer2009scalable} or both \cite{jacob2011jackstraws}. For instance, Andronio \cite{andronio2015heldroid} proposes an approach that identifies device-locking or encryption activities at the system level by finding code paths using static taint analysis along with symbolic execution. In another work, anomalous file system activities are used to detect ransomware \cite{kharraz2015cutting}. Similarly, Scaife \etal \cite{scaife2016cryptolock} attempts to identify abnormal system behaviour by carefully measuring changes in file type, similarity measurements and entropy.
\begin{figure}[t!]
	\centering
	\includegraphics[width=0.99\linewidth]{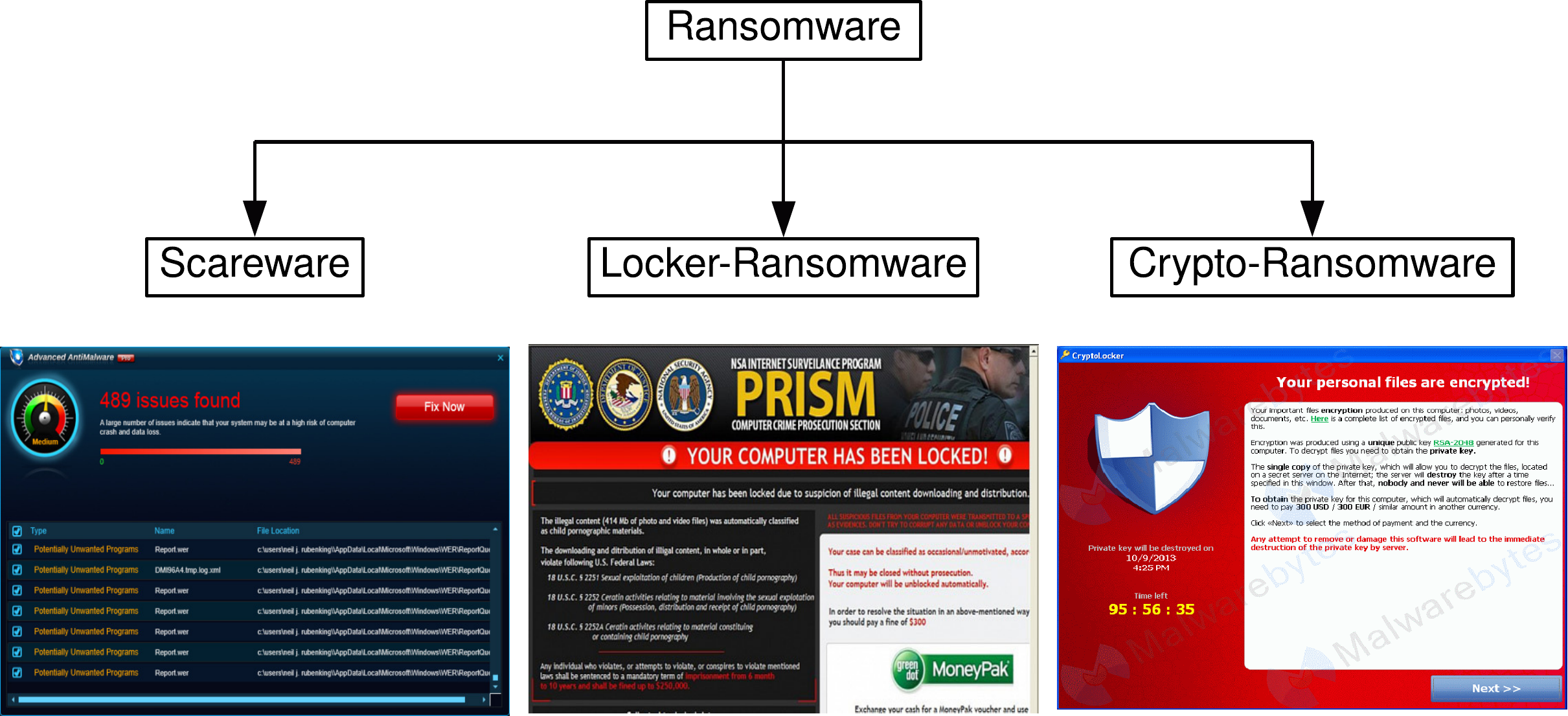}
	\captionsetup[figure]{skip=7pt}
	\captionof{figure}{Types of ransomware grouped by attack intensity.}%
	\label{fig:ranomware_tax}\vspace{-0.6cm}
\end{figure}

With\blfootnote{\textit{Volenti non fit injuria}: No wrong is done to one who consents.} the significant recent advances in machine learning \cite{simonyan2014very, atapour2018real, ren2015faster, mikolov2013distributed, grover2016node2vec, bonner2018temporal}, learning-based approaches have also found their way into the expansive literature on ransomware detection and classification. For example, the approach proposed by Sgandurra \etal \cite{sgandurra2016automated} detects and classifies ransomware variants by dynamically analysing the behaviour of applications during the early stages of their installation. Ransomware classification has also been attempted through combining a static detection phase based on the frequency of opcodes prior to installation and a dynamic method which investigates the use of CPU, memory and network as well as call statistics during run-time \cite{ferrante2017extinguishing}. Vinayakumar \etal \cite{vinayakumar2017evaluating} investigates the efficacy of neural networks used to detect and classify ransomware activities, with a focus on tuning the hyperparameters and the architecture of a simple multilayer perceptron. In another work, Atapour \etal \cite{aaa} propose a vision-based system that classifies ransomware variants based on an image of the splash screen casually captured using a smartphone camera.

Despite the advances in ransomware detection techniques, the constantly-evolving landscape of ransomware and the substantial level of diversity among its variants give further importance to acquiring deeper insight into the nature of ransomware attacks. In this vein, we take a closer look at ransomware categories and particularly the types of victims that are often targeted and regularly fall for ransomware attacks. In Section \ref{sec:ransomware}, we briefly review the types of ransomware commonly found in the wild, Section \ref{sec:victims} focuses on the victims of ransomware attacks and finally, a number of preventative and response strategies are discussed in Section \ref{sec:prevention_response}.

\section{Variants of Ransomware}
\label{sec:ransomware}

Before attempting to understand the victims of ransomware attacks, it is important to understand the varieties of ransomware and the reasons behind this variation. While the existing literature contains numerous studies that provide meaningful taxonomies of ransomware \cite{luo2007awareness, luo2009ransomware,ahmadian2015connection, al2018ransomware}, we primarily focus on aspects of ransomware variants that can directly contribute to a deeper understanding of their victims.

Ransomware variants can be classified based on their mode of propagation (\eg through pre-packaged exploitation kits \cite{hopkins2015exploit}, affiliate packages built on top of existing malware infrastructures \cite{wyke2012zeroaccess}, spam campaigns \cite{wyke2012zeroaccess}), payment methods (\eg direct digital currency payments \cite{coindesk}, pre-paid vouchers, calls and texts to premium rate numbers, online purchases) and many other characteristics. However, of the numerous factors that can aid in the classification of ransomware with respect to the victims: attack intensity \cite{luo2007awareness, luo2009ransomware} and the user platform towards which the attack is designed \cite{al2018ransomware} are arguably the most important. Considering these factors, we provide a very generic classification of ransomware variants in the following.

\subsection{Attack Intensity}
\label{ssec:intensity}

In terms of the intensity of the attacks (the level of threat a ransomware can pose to an infected system), there are, in general, three types of ransomware: scareware, locker-ransomware and crypto-ransomware (Figure \ref{fig:ranomware_tax}).

Predominantly, the objective of a scareware is simply to scare the victim into paying a fee without causing any actual harm to the computer system \cite{savage} and can thus be easily dealt with. This is usually accomplished by displaying a fake splash screen on the victim's computer \cite{aaa} and asking for a ransom despite the files and the entire system still being accessible to the user. In most cases, a scareware might threaten the victim by alleging that they have found illegal content or viral infections on the system \cite{richet2016extortion, pathak2016dangerous} and exploit the victim's fear to extort money.

Conversely, locker-ransomware and crypto-ransomware \cite{cabaj2018software} can be truly detrimental to system, sometimes causing irreparable damage. Locker-ransomware often takes control of the locking capabilities of the operating system and denies the victim access to one or more of the system services or applications \cite{savage}. The system is subsequently left with limited capabilities, which might only allow the victim to follow the instructions needed to pay the ransom. However, this form of ransomware is, in many cases, incapable of successfully extorting the ransom, especially from skilled PC users, as the operating system and the files are left intact and unharmed and it is relatively easy to bypass the locking mechanism.

Crypto-ransomware, on the other hand, employs cryptography to encrypt the user's files \cite{ganesh2016static}, essentially removing all access to the files, leaving the victim with two options: either pay the ransom or forever lose access to all the encrypted files (which in many cases will happen even if the ransom is paid). In essence, the ransom is demanded in exchange for the decryption key, which is often the only way for the victim to regain access to the data and/or the system.

When it comes to the profile of the victims targeted by these types of ransomware, technical knowledge often plays a critical role. While highly-skilled victims are unlikely to pay the ransom, except for cases where a crypto-ransomware has successfully encrypted files which have not been archived or backed up, unskilled users can fall victim to locker-ransomware or even scareware. However, depending on the platform, these variants of ransomware can have different effects and can cause varying levels of harm. As such, in the following section, we focus on the platforms different variants of ransomware might target.

\subsection{Target Platform}
\label{ssec:platform}

Another factor that plays a role in understanding the ransomware victims is the platform they use. While PCs and mobile devices have long been the target of ransomware attacks, new variants of ransomware now commonly attack IoT devices and cloud-based systems as well \cite{symantec2019internet}. There has even been a demonstration that other consumer devices such as digital cameras can also be successfully targeted \cite{eos}. Encrypting images directly on the camera could have significant negative impact, especially if the photographer is working as a professional.

However, due to the wide-spread use of personal computers for decades, it is expected that they make up the majority of ransomware targets, and while MS Windows systems are most commonly attacked, others such as Mac OS and Linux machines are not entirely immune either \cite{arsene2016ransomware, benchea2016}. PCs are often targeted by all three types of ransomware (scareware, locker-ransomware and crypto-ransomware), with the number of attacks being on the rise and new variants constantly being introduced \cite{Symantec, mcafee2017mcafee}.

Due to the ease of use and the low skill levels needed to operate, mobile devices are now ubiquitously used by a wide spectrum of individuals, making them ideal targets for ransomware attacks \cite{yang2015automated, afifi2016dyhap}. Mobile ransomware attacks have reportedly more than quadrupled since 2015 \cite{ksn}, with locker-ransomware variants carrying out most of the successful attacks. This is mainly due to the fact that important personal files are often kept outside the mobile device, rendering local encryption attacks against the device useless, and the mobile operating systems do not offer the manoeuvrability needed to bypass a locking attack, making them significantly more effective \cite{al2018ransomware} than locking attacks against a PC.

Recently, there have been numerous reports of attacks on IoT devices \cite{karkouch2016data}, despite such appliances generally not holding any valuable data. Not unlike mobile attacks, different variants of locker-ransomware can inflict significant harm to the users of IoT devices by disabling access, causing power outages and even disrupting critical services \cite{Symantec}. To understand why and how certain victims are targeted more often than others, in the next section, we focus on the victims, themselves.

\section{Ransomware Victims}
\label{sec:victims}

Since significant financial gain continuously drives the creation and spread of ransomware \cite{laszka2017economics}, one of the most effective methods of combating this type of malware is cutting off the supply of funds obtained through the ransom paid by the victims. Hence, a deeper understanding of the typical victims of ransomware attacks can be very helpful in coming up with solutions that prevent or mitigate the current fast-growing ransomware problem. From a top-down perspective, ransomware victims can broadly be classified into two wide groups: individuals and business entities. Due to the important differences between these two groups, the behaviour of the ransomware targeting these victims is often very different.

\subsection{Individual Home Users}
\label{ssec:individuals}

Consumer ransomware often targets individual home users, which in terms of numbers make up the majority of ransomware victims \cite{al2018ransomware}. These individualistic attacks are often opportunistic and perpetrated via indiscriminate attack vectors. For instance, the victim might receive a spam e-mail, in which they are encouraged to click on a malicious link or they might visit a compromised website infecting the system with ransomware. In rarer occasions, however, infection can occur without user engagement through drive-by downloads \cite{cova2010detection} or by means of malvertising and ad-injections \cite{xing2015understanding}.

Considering the limited resources often available to non-technical individual victims compared to large corporations and government-affiliated organisations, the ransom demanded from the victims is often significantly smaller (\$300 to \$700) \cite{Symantec}. When a consumer ransomware attacks a targeted individual, all the files and resources are normally locked or encrypted as fast as computationally possible and the ransom note is quickly displayed in the form of a splash screen \cite{aaa}. Despite the more affordable fees demanded by the perpetrators, due to the large number of infections a single attack can spread across the world, this type of ransomware attack remains profitable and incentivises further investment of resources and development for the perpetrators \cite{bisson2017half}.

Non-technical individuals are also widely targeted by scareware (Section \ref{ssec:intensity}), the variants of which essentially issue fake warnings and threaten the victim's files and/or personal privacy without them ever actually being in any serious danger. For instance, the famous FakeAV \cite{stone2013underground} epitomises a typical scareware by adopting the appearance of a legitimate anti-virus that warns the user of the supposedly malicious software it has discovered on the victim's computer after a fake scan. Subsequent to the this, payment is demanded, sometimes very aggressively, to remove the fake malware \cite{savage}.

With the growth of the \emph{ransomware-as-a-service} model \cite{nadir2018contemporary}, even unskilled amateur hackers are now capable of launching ransomware attacks using pre-fabricated automated tools, which sometimes come with a consumer support service to talk the victims through negotiation attempts and payments \cite{sherer2016ransonware}. This has a led to an increase in the number of mass indiscriminate attacks against individuals, necessitating a deeper analysis and understanding of the situation. While certain factors such as age, level of education and financial resources do contribute to the likelihood of an individual falling victim to a ransomware attack and subsequently paying the ransom, the level of computer-literacy is the primary determining factor.

While simple solutions such as regular software and operating system updates \cite{savage, leong2016understanding}, e-mail security (on both client and server side \cite{cormack2008email, leong2016understanding}), anti-malware tools \cite{continella2016shieldfs}, access and authorisation control \cite{mattei2017privacy} and simply backing up the data \cite{young2017cryptovirology, mustaca2014your} can significantly reduce the number of successful ransomware attacks on individuals, the user needs to be aware of and skilled enough to implement such measures, which signifies the value of computer-literacy for the public.

\subsection{Business Entities}
\label{ssec:business}

Despite having access to large IT infrastructures and security professionals, business organisations also regularly fall victim to ransomware attacks. Such ransomware attacks are often a consequence of the more targeted (big game hunting) attacks where the perpetrator may put significant effort into preparing a credible social attack against an identified member of staff. After gaining access through an entry point into the system, the attack vectors often employed by such ransomware variants are mostly gradual and covert \cite{Symantec}. The ransomware usually focuses on avoiding and evading the countermeasures deployed by the organisation's security experts and slowly takes control of specifically targeted data, such as transactional documents, backups and archives \cite{al2018ransomware}. As expected, large organisations often receive significantly higher ransom demands compared to individuals, easily reaching numbers as high as \$10,000 or higher \cite{al2018ransomware}.

As far as business entities and organisations are concerned, the security systems, the type of data and possibly the services they deal with are the primary factors in their victimisation. In the following, we focus on the various sectors that are often targeted by ransomware attacks.

\subsubsection{Education}
\label{sssec:education}

In recent years, educational institutions, such as schools and universities, have become one of the primary targets of ransomware attacks. In fact, according to a recent report \cite{bitsite}, the educational sector faces the largest number of attacks per capita with more than 10\% of all schools and universities having been targeted. Many such organisations have budgetary constraints, limited access to cybersecurity professionals and smaller teams of often over-worked IT personnel, yet due to their high rate of network file sharing and centralised systems \cite{target} can be prime targets for any malware.

Additionally, an average school, university or any other educational or research institute stores valuable and highly-sensitive data on students, who might be under the legal age, staff, intellectual property, financial documents and sometimes even medical records that must not be compromised in any way. One such ransomware attack was experienced by University College London, where shared drives and student management systems were compromised \cite{ucl} in 2017.

\subsubsection{Health Industry}
\label{sssec:health}

Healthcare facilities and hospitals are also commonly targeted for ransomware attacks, mainly due to the highly critical data and services that they depend on. Not having immediate access to a patient's data can have life-or-death consequences. A major example of this would be the Hollywood Presbyterian Medical Center, where a ransom of \$3.7 million was demanded after highly-sensitive medical data and hospital services were disrupted. The hospital was forced to pay the ransom since daily administrative operations were reduced to pen and paper, and more importantly, the life of the patients was hanging in the balance. This case, however, is particularly notable as the ransom was negotaited down to \$17,000 and access to the data and services was returned to the hospital \cite{mattei2017privacy} allowing normal operations to resume.

\subsubsection{Government Agencies}
\label{sssec:gov}

The number of attacks on government agencies tripled from 2015 to 2016 \cite{bitsite} and has been steadily growing ever since. In 2018, the city of Atlanta, Georgia suffered a significant ransomware attack. The ransomware had found its entry point into the system by means of a brute-force attack to crack weaker passwords \cite{statescoop}. Although most services (safety, water and airport operations) were not compromised, online payment systems and court information access were severely restricted, potentially affecting up to 6 million people \cite{statescoop} and costing the city over \$2.7 million in recovery efforts.

As government agencies are generally perceived to have significantly larger funds available to them and many of their services (\eg police, water, transportation) are highly critical and time-sensitive, perpetrators always look for opportunities to get through the security systems of such organisations in any way possible.

\subsubsection{Utilities, Retail and Finance}
\label{sssec:others}

With the growing reliance many organisations in the utilities, retail and finance sector place on computationally-powerful low-latency systems, combined with the significantly heavy costs they can suffer as a result of any down-time, they have become one of the recurring targets of ransomware attacks \cite{bitsite}. In many such companies, and even small to mid-size businesses of similar nature, the human resources departments are now regularly preyed on as they often have access to many other sections and departments within any given company and this connectivity is very enticing to the perpetrators \cite{csoonline}.

\subsubsection{Emerging Targets}
\label{sssec:emerging}

In general, any organisation that holds sensitive data or offers critical services is always at risk of a ransomware attack. A law firm, for instance, is always in danger. While the loss of data can be catastrophic to a law firm, the possibility of publicising confidential client data can put an end to the business entirely, which would make a law firm willing to pay any amount \cite{target}. Industrial control systems have also largely avoided being targets of ransomware attacks \cite{formby2017out}, but this is not because of their level of security as there are glaring security vulnerabilities that do not seem to be improving \cite{savage}. Any widespread attack on such systems can lead to a compromise in critical infrastructure, which can have devastating international consequences.

\section{Prevention and Response}
\label{sec:prevention_response}

Since the significant re-emergence of ransomware within the past few years, many new tools and workarounds have been suggested to mitigate or recover from a ransomware attack. However, due to the increasing viability of the business model, perpetrators invest significant resources to stay ahead of the cybersecurity community. According to a recent report \cite{nahorney2017internet}, the number of ransomware attacks that have been successfully detected and prevented saw a 30\% increase from 2015 to 2016. However, the overall number of attacks has also notably risen. Consequently, detecting every new variant of ransomware might be impossible using a single powerful anti-malware tool, but there are various techniques that individuals or companies can employ to protect their systems from a ransomware infection. Here, we discuss certain measures that can help with the prevention of or response to a ransomware attack. In Section \ref{ssec:prevention}, we discuss some of the most prominent security techniques that can prevent ransomware attacks and in Section \ref{ssec:response}, we focus on what should be done in response to an infection.

\subsection{Prevention Techniques}
\label{ssec:prevention}

Attempting to remove a ransomware or re-gaining access to a corrupted system or encrypted data can be very expensive and sometimes impossible, even if the ransom is paid. The best solution, in this case, is prevention. While securing the system and network activity is extremely important for both individual users and business organisations, an overwhelming majority of successful malware attacks are due to human error \cite{proofpoint}, so securing the end user is of utmost importance. In the following, we briefly outline the most important prevention approaches on the system and the user side.

\subsubsection{Network/System Security}
\label{sssec:network-system}

While certain system and network security measures are more expensive and require expert support, the majority of the recommendations listed below apply to both individual home users and business entities:

\begin{itemize}
	\item A robust backup and archiving system removes the threat the majority of ransomware attacks can pose.
	\item Anti-malware and other similar tools are indispensable to a secure system. Modern tools \cite{continella2016shieldfs} even take advantage of machine learning approaches to remove their dependence on knowledge of existing threat signatures.
	\item In a secure system, all hardware, operating systems, software, cloud locations and content management systems must be patched and up-to-date at all times.
	\item Via effective system administration, application white-listing and software restriction policies \cite{sittig2016socio}, dubious programs can be kept off the system, specially for large companies, where controlling and monitoring all the employees with system access might not be possible.
	\item Using a proxy-server and any of the numerous ad-blocking packages, common ransomware entry points can be restricted.
	\item Through network segmentation, virtual machines, and limited authorisation and privileges, potentially harmful network access to sensitive data can be averted.
	\item It is important for any organisation, to introduce access policies and closely monitor any third parties as they can easily introduce vulnerabilities into the system.
	\item Any company with sensitive data requires a response plan that outlines how the system can be protected if an attack is detected in its early stages.
\end{itemize}

\subsubsection{End User Security}
\label{sssec:user}

While human error is a significant cause of ransomware attacks \cite{proofpoint}, it is not always avoidable. Nor can we assume that a well-constructed attack would not thwart even the most security-savvy. Education and training, both for individuals and businesses with many employees, can be very effective in preventing, or at least reducing the likelihood of, a ransomware attack.

\begin{itemize}
	\item End users need to be aware of social engineering as it is a common attack vector for many malicious actors \cite{gallegos2017social}.
	\item All end users must be trained on phishing and how it must be countered.
	\item Companies should have strict policies about their employees' use of the internet as personal emails and social media websites \cite{richardson2017ransomware} are regular points of entry for many ransomware variants.
	\item It is very important for end users to have strong passwords as perpetrators can easily gain access to the main system via various password cracking approaches  \cite{marechal2008advances}.
\end{itemize}

\subsection{Response to an Attack}
\label{ssec:response}

More often than not, when data has been encrypted using a crypto-ransomware, not many solutions are left available. However, it is very important, especially for non-skilled individuals, to ensure the infection has indeed been caused by a crypto-ransomware as locker-ransomware and scareware can often be easily removed from most computer systems without causing serious harm to the system.

Identification is often key when seeking to distinguish which type of Ransomware has infected the system. In such situations, knowledge is highly-important especially as many attackers obfuscate their ransomware to reduce the chance of easy detection. The \enquote{No More Ransom} project \cite{nomoreransom} provides a mechanism to identify the ransomware from either the text within the ransom note or a small number of the encrypted files. While Atapour \etal offers a more layperson approach allowing users to take a picture of the ransomware splash screen and use this for identification~\cite{aaa}.

In cases where the algorithm or the key used by the perpetrators are not strong, a decryption solution might be available. While certain companies such as Kaspersky and Windows Defender offer proprietary decryption tools, the \emph{No More Ransom} project \cite{nomoreransom} is specifically dedicated to helping all victims, whether individual home users or businesses, to recover their encrypted files without having to pay the ransom.

Finally, one of the most important actions every business entity or individual home user must take after a ransomware infection is to report the incident and share as much data about the incident as possible with the authorities and experts. Many individuals and companies often remain quiet about such attacks out of fear of bad publicity. However, reporting such events can go a long way in putting a stop to similar future attacks.

\section{Conclusion}

The level of threat that ransomware poses is extremely serious and can disrupt the fabric of our modern data-dependent society. With the recent rise in cybercrime, it is now more important than ever to combat the perpetrators of ransomware attacks and cut off the large supply of funds regularly invested in the development and improvement of new variants of ransomware. In this vein, this paper has primarily focused on facilitating a better understanding of ransomware variants and the victims often targeted by them. We also provide a brief classification of ransomware variants and the attack vectors they are commonly associated with. This is accomplished by examining the severity of the threat a variant of ransomware can pose and the platform it is designed to attack. The paper also discusses the targets predominantly victimised by ransomware perpetrators to enable further insight into the underlying forces that drive this malicious business model. While individual home users make up the majority of the victims, depending on the type of data they hold or the services they may provide, businesses can make for more lucrative targets and are often at a greater risk. Furthermore, we have briefly considered helpful prevention strategies for individuals and businesses that fall victim to these attacks and the potential post-infection recovery techniques that might aid in mitigating the devastating effects the loss of sensitive data can have on any victim as paying the ransom is rarely advisable and is not guaranteed to lead to a full recovery of the data.


\section*{Acknowledgement}

This work was in part supported by the EPSRC EMPHASIS (EP/P01187X/1) and CRITiCaL (EP/M020576/1) projects.


\bibliographystyle{IEEEtran}
\bibliography{ref}

\begin{thebibliography}{10}
\providecommand{\url}[1]{#1}
\csname url@samestyle\endcsname
\providecommand{\newblock}{\relax}
\providecommand{\bibinfo}[2]{#2}
\providecommand{\BIBentrySTDinterwordspacing}{\spaceskip=0pt\relax}
\providecommand{\BIBentryALTinterwordstretchfactor}{4}
\providecommand{\BIBentryALTinterwordspacing}{\spaceskip=\fontdimen2\font plus
\BIBentryALTinterwordstretchfactor\fontdimen3\font minus
  \fontdimen4\font\relax}
\providecommand{\BIBforeignlanguage}[2]{{%
\expandafter\ifx\csname l@#1\endcsname\relax
\typeout{** WARNING: IEEEtran.bst: No hyphenation pattern has been}%
\typeout{** loaded for the language `#1'. Using the pattern for}%
\typeout{** the default language instead.}%
\else
\language=\csname l@#1\endcsname
\fi
#2}}
\providecommand{\BIBdecl}{\relax}
\BIBdecl

\bibitem{zhang2019classification}
H.~Zhang, X.~Xiao, F.~Mercaldo, S.~Ni, F.~Martinelli, and A.~K. Sangaiah,
  ``Classification of ransomware families with machine learning based on
  {N}-gram of opcodes,'' \emph{Future Generation Computer Systems}, vol.~90,
  pp. 211--221, 2019.

\bibitem{moore2016detecting}
C.~Moore, ``Detecting ransomware with honeypot techniques,'' in
  \emph{Cybersecurity and Cyberforensics Conference}.\hskip 1em plus 0.5em
  minus 0.4em\relax IEEE, 2016, pp. 77--81.

\bibitem{laszka2017economics}
A.~Laszka, S.~Farhang, and J.~Grossklags, ``On the economics of ransomware,''
  in \emph{Int. Conf. Decision and Game Theory for Security}.\hskip 1em plus
  0.5em minus 0.4em\relax Springer, 2017, pp. 397--417.

\bibitem{kok2019ransomware}
S.~Kok, A.~Abdullah, N.~Jhanjhi, and M.~Supramaniam, ``Ransomware, threat and
  detection techniques: A review,'' \emph{Int. J. Computer Science and Network
  Security}, vol.~19, no.~2, p. 136, 2019.

\bibitem{fbi}
``High-impact ransomware attacks threaten {U.S.} businesses and
  organisations,'' https://www.ic3.gov/media/2019/191002.aspx.

\bibitem{gu2008botminer}
G.~Gu, R.~Perdisci, J.~Zhang, and W.~Lee, ``Botminer: Clustering analysis of
  network traffic for protocol and structure independent {Botnet} detection,''
  \emph{USENIX Security Symposium}, 2008.

\bibitem{cabaj2018software}
K.~Cabaj, M.~Gregorczyk, and W.~Mazurczyk, ``Software-defined networking-based
  crypto ransomware detection using {HTTP} traffic characteristics,''
  \emph{Computers \& Electrical Engineering}, vol.~66, pp. 353--368, 2018.

\bibitem{bayer2009scalable}
U.~Bayer, P.~M. Comparetti, C.~Hlauschek, C.~Kruegel, and E.~Kirda, ``Scalable,
  behavior-based malware clustering,'' in \emph{Network and Distributed System
  Security Symposium}, vol.~9, 2009, pp. 8--11.

\bibitem{jacob2011jackstraws}
G.~Jacob, R.~Hund, C.~Kruegel, and T.~Holz, ``{JACKSTRAWS}: Picking command and
  control connections from {Bot} traffic,'' in \emph{USENIX Security
  Symposium}, 2011.

\bibitem{andronio2015heldroid}
N.~Andronio, ``Heldroid: Fast and efficient linguistic-based ransomware
  detection,'' Ph.D. dissertation, 2015.

\bibitem{kharraz2015cutting}
A.~Kharraz, W.~Robertson, D.~Balzarotti, L.~Bilge, and E.~Kirda, ``Cutting the
  gordian knot: {A} look under the hood of ransomware attacks,'' in \emph{Int.
  Conf. Detection of Intrusions and Malware, and Vulnerability
  Assessment}.\hskip 1em plus 0.5em minus 0.4em\relax Springer, 2015, pp.
  3--24.

\bibitem{scaife2016cryptolock}
N.~Scaife, H.~Carter, P.~Traynor, and K.~R. Butler, ``Cryptolock (and drop it):
  Stopping ransomware attacks on user data,'' in \emph{Int. Conf. Distributed
  Computing Systems}.\hskip 1em plus 0.5em minus 0.4em\relax IEEE, 2016, pp.
  303--312.

\bibitem{simonyan2014very}
K.~Simonyan and A.~Zisserman, ``Very deep convolutional networks for
  large-scale image recognition,'' \emph{arXiv preprint arXiv:1409.1556}, 2014.

\bibitem{atapour2018real}
A.~Atapour-Abarghouei and T.~P. Breckon, ``Real-time monocular depth estimation
  using synthetic data with domain adaptation via image style transfer,'' in
  \emph{IEEE Conf. Computer Vision and Pattern Recognition}, 2018, pp.
  2800--2810.

\bibitem{ren2015faster}
S.~Ren, K.~He, R.~Girshick, and J.~Sun, ``Faster {R-CNN}: Towards real-time
  object detection with region proposal networks,'' in \emph{Advances in Neural
  Information Processing Systems}, 2015, pp. 91--99.

\bibitem{mikolov2013distributed}
T.~Mikolov, I.~Sutskever, K.~Chen, G.~S. Corrado, and J.~Dean, ``Distributed
  representations of words and phrases and their compositionality,'' in
  \emph{Advances in Neural Information Processing Systems}, 2013, pp.
  3111--3119.

\bibitem{grover2016node2vec}
A.~Grover and J.~Leskovec, ``{node2vec}: Scalable feature learning for
  networks,'' in \emph{Int. Conf. Knowledge Discovery and Data Mining}.\hskip
  1em plus 0.5em minus 0.4em\relax ACM, 2016, pp. 855--864.

\bibitem{bonner2018temporal}
S.~Bonner, J.~Brennan, I.~Kureshi, G.~Theodoropoulos, A.~S. McGough, and
  B.~Obara, ``Temporal graph offset reconstruction: Towards temporally robust
  graph representation learning,'' in \emph{IEEE Int. Conf. Big Data}, 2018,
  pp. 3737--3746.

\bibitem{sgandurra2016automated}
D.~Sgandurra, L.~Mu{\~n}oz-Gonz{\'a}lez, R.~Mohsen, and E.~C. Lupu, ``Automated
  dynamic analysis of ransomware: Benefits, limitations and use for
  detection,'' \emph{arXiv preprint arXiv:1609.03020}, 2016.

\bibitem{ferrante2017extinguishing}
A.~Ferrante, M.~Malek, F.~Martinelli, F.~Mercaldo, and J.~Milosevic,
  ``Extinguishing ransomware - {A} hybrid approach to {Android} ransomware
  detection,'' in \emph{Int. Symp. Foundations and Practice of Security}.\hskip
  1em plus 0.5em minus 0.4em\relax Springer, 2017, pp. 242--258.

\bibitem{vinayakumar2017evaluating}
R.~Vinayakumar, K.~Soman, K.~S. Velan, and S.~Ganorkar, ``Evaluating shallow
  and deep networks for ransomware detection and classification,'' in
  \emph{Int. Conf. Advances in Computing, Communications and
  Informatics}.\hskip 1em plus 0.5em minus 0.4em\relax IEEE, 2017, pp.
  259--265.

\bibitem{aaa}
A.~Atapour-Abarghouei, S.~Bonner, and A.~S. McGough, ``A king’s ransom for
  encryption: Ransomware classification using augmented one-shot learning and
  bayesian approximation,'' in \emph{IEEE Int. Conf. Big Data}, 2019, pp. 1--6.

\bibitem{luo2007awareness}
X.~Luo and Q.~Liao, ``Awareness education as the key to ransomware
  prevention,'' \emph{Information Systems Security}, vol.~16, no.~4, pp.
  195--202, 2007.

\bibitem{luo2009ransomware}
------, ``Ransomware: A new cyber hijacking threat to enterprises,'' in
  \emph{Handbook of Research on Information Security and Assurance}.\hskip 1em
  plus 0.5em minus 0.4em\relax IGI global, 2009, pp. 1--6.

\bibitem{ahmadian2015connection}
M.~M. Ahmadian, H.~R. Shahriari, and S.~M. Ghaffarian, ``Connection-monitor \&
  connection-breaker: A novel approach for prevention and detection of high
  survivable ransomwares,'' in \emph{Int. Iranian Society of Cryptology Conf.
  Information Security and Cryptology}.\hskip 1em plus 0.5em minus 0.4em\relax
  IEEE, 2015, pp. 79--84.

\bibitem{al2018ransomware}
B.~A.~S. Al-rimy, M.~A. Maarof, and S.~Z.~M. Shaid, ``Ransomware threat success
  factors, taxonomy, and countermeasures: A survey and research directions,''
  \emph{Computers \& Security}, vol.~74, pp. 144--166, 2018.

\bibitem{hopkins2015exploit}
M.~Hopkins and A.~Dehghantanha, ``Exploit {Kits}: The production line of the
  cybercrime economy?'' in \emph{Int. Conf. Information Security and Cyber
  Forensics}.\hskip 1em plus 0.5em minus 0.4em\relax IEEE, 2015, pp. 23--27.

\bibitem{wyke2012zeroaccess}
J.~Wyke, ``The {ZeroAccess} {Botnet--Mining} and fraud for massive financial
  gain,'' \emph{Sophos Technical Paper}, 2012.

\bibitem{coindesk}
S.~Higgins, ``{CoinDesk} on {Citrix} survey, {BTC} stocks in {UK} businesses,''
  https://www.coindesk.com/survey-uk-bitcoin-ransomware.

\bibitem{savage}
K.~Savage, P.~Coogan, and H.~Lau, ``The evolution of ransomware,''
  \emph{Security Response: Symantec Corporation}, 2015.

\bibitem{richet2016extortion}
J.-L. Richet, ``Extortion on the internet: the rise of crypto-ransomware,''
  \emph{Harvard}, 2016.

\bibitem{pathak2016dangerous}
P.~Pathak and Y.~M. Nanded, ``A dangerous trend of cybercrime: Ransomware
  growing challenge,'' \emph{Int. J. Advanced Research in Computer Engineering
  \& Technology}, vol.~5, no.~2, pp. 371--373, 2016.

\bibitem{ganesh2016static}
N.~Ganesh, F.~Di~Troia, V.~A. Corrado, T.~H. Austin, and M.~Stamp, ``Static
  analysis of malicious {Java} applets,'' in \emph{Int. Workshop on Security
  And Privacy Analytics}.\hskip 1em plus 0.5em minus 0.4em\relax ACM, 2016, pp.
  58--63.

\bibitem{symantec2019internet}
Symantec, ``Internet security threat report,'' 2019.

\bibitem{eos}
``Say cheese, ransomware-ing a {DSLR} camera,''
  https://research.checkpoint.com/say-cheese-ransomware-ing-a-dslr-camera/.

\bibitem{arsene2016ransomware}
L.~Arsene and A.~Gheorghe, ``Ransomware: {A} victim’s perspective,''
  \emph{BitDefender}, 2016.

\bibitem{benchea2016}
R.~Benchea, V.~Cristina, M.~Alexandru, and L.~Arsene, ``Petya ransomware goes
  low level,'' in \emph{BitDefender}.\hskip 1em plus 0.5em minus 0.4em\relax
  BitDefender, 2016.

\bibitem{Symantec}
J.-P. P., D.~OBrien, and S.~Wallace, ``Ransomware and businesses,'' \emph{An
  ISTR Special Report: Symantec Corporation}, 2016.

\bibitem{mcafee2017mcafee}
McAfee, ``Threats predictions,'' \emph{McAfee LLC.}, 2017.

\bibitem{yang2015automated}
T.~Yang, Y.~Yang, K.~Qian, D.~C.-T. Lo, Y.~Qian, and L.~Tao, ``Automated
  detection and analysis for {Android} ransomware,'' in \emph{Int. Conf. High
  Performance Computing and Communications, Int. Symp. Cyberspace Safety and
  Security, and Int. Conf. Embedded Software and Systems}.\hskip 1em plus 0.5em
  minus 0.4em\relax IEEE, 2015, pp. 1338--1343.

\bibitem{afifi2016dyhap}
F.~Afifi, N.~B. Anuar, S.~Shamshirband, and K.-K.~R. Choo, ``{DyHAP}: Dynamic
  hybrid {ANFIS-PSO} approach for predicting mobile malware,'' \emph{PloS one},
  vol.~11, no.~9, p. e0162627, 2016.

\bibitem{ksn}
K.~Lab, ``{KSN} report: Ransomware,'' 2016.

\bibitem{karkouch2016data}
A.~Karkouch, H.~Mousannif, H.~Al~Moatassime, and T.~Noel, ``Data quality in
  internet of things: {A} state-of-the-art ssurvey,'' \emph{Journal of Network
  and Computer Applications}, vol.~73, pp. 57--81, 2016.

\bibitem{cova2010detection}
M.~Cova, C.~Kruegel, and G.~Vigna, ``Detection and analysis of
  drive-by-download attacks and malicious {JavaScript} code,'' in
  \emph{International Conference World Wide Web}.\hskip 1em plus 0.5em minus
  0.4em\relax ACM, 2010, pp. 281--290.

\bibitem{xing2015understanding}
X.~Xing, W.~Meng, B.~Lee, U.~Weinsberg, A.~Sheth, R.~Perdisci, and W.~Lee,
  ``Understanding malvertising through ad-injecting browser extensions,'' in
  \emph{Int. Conf. World Wide Web}.\hskip 1em plus 0.5em minus 0.4em\relax
  International World Wide Web Conferences Steering Committee, 2015, pp.
  1286--1295.

\bibitem{bisson2017half}
D.~Bisson, ``Half of {American} ransomware victims have paid the ransom,''
  \emph{TripWire}, 2017.

\bibitem{stone2013underground}
B.~Stone-Gross, R.~Abman, R.~A. Kemmerer, C.~Kruegel, D.~G. Steigerwald, and
  G.~Vigna, ``The underground economy of fake antivirus software,'' in
  \emph{Economics of Information Security and Privacy III}.\hskip 1em plus
  0.5em minus 0.4em\relax Springer, 2013, pp. 55--78.

\bibitem{nadir2018contemporary}
I.~Nadir and T.~Bakhshi, ``Contemporary cybercrime: A taxonomy of ransomware
  threats \& mitigation techniques,'' in \emph{Int. Conf. Computing,
  Mathematics and Engineering Technologies}.\hskip 1em plus 0.5em minus
  0.4em\relax IEEE, 2018, pp. 1--7.

\bibitem{sherer2016ransonware}
J.~A. Sherer, M.~L. McLellan, E.~R. Fedeles, and N.~L. Sterling,
  ``Ransonware-practical and legal considerations for confronting the new
  economic engine of the dark web,'' \emph{Rich. JL \& Tech.}, vol.~23, p.~1,
  2016.

\bibitem{leong2016understanding}
R.~Leong, C.~Beek, C.~Cochin, N.~Cowie, and C.~Schmugar, ``Understanding
  ransomware and strategies to defeat it,'' \emph{White Paper (McAfee Labs)},
  pp. 1--16, 2016.

\bibitem{cormack2008email}
G.~V. Cormack \emph{et~al.}, ``Email spam filtering: {A} systematic review,''
  \emph{Foundations and Trends{\textregistered} in Information Retrieval},
  vol.~1, no.~4, pp. 335--455, 2008.

\bibitem{continella2016shieldfs}
A.~Continella, A.~Guagnelli, G.~Zingaro, G.~De~Pasquale, A.~Barenghi,
  S.~Zanero, and F.~Maggi, ``{ShieldFS}: {A} self-healing, ransomware-aware
  filesystem,'' in \emph{Annual Conf. Computer Security Applications}.\hskip
  1em plus 0.5em minus 0.4em\relax ACM, 2016, pp. 336--347.

\bibitem{mattei2017privacy}
T.~A. Mattei, ``Privacy, confidentiality, and security of health care
  information: Lessons from the recent {Wannacry} cyberattack,'' \emph{World
  Neurosurgery}, vol. 104, pp. 972--974, 2017.

\bibitem{young2017cryptovirology}
A.~L. Young and M.~Yung, ``Cryptovirology: The birth, neglect, and explosion of
  ransomware,'' \emph{Communications of the ACM}, vol.~60, no.~7, pp. 24--26,
  2017.

\bibitem{mustaca2014your}
S.~Mustaca, ``Are your {IT} professionals prepared for the challenges to
  come?'' \emph{Computer Fraud \& Security}, vol. 2014, no.~3, pp. 18--20,
  2014.

\bibitem{bitsite}
N.~Simon, ``The rising face of cybercrime: Ransomware,''
  https://www.bitsight.com/blog/rising-face-of-cybercrime-ransomware.

\bibitem{target}
J.~Martin, ``Who is a target for ransomware attacks?''
  https://www.csoonline.com/article/3208111/who-is-a-target-for-ransomware-attacks.html.

\bibitem{ucl}
A.~Hern, ``{University College London} hit by ransomware attack,''
  https://www.theguardian.com/technology/2017/jun/15/university-college-london-hit-by-ransomware-attack-hospitals-email-phishing.

\bibitem{statescoop}
B.~Freed, ``Atlanta was not prepared to respond to a ransomware attack,''
  https://statescoop.com/atlanta-was-not-prepared-to-respond-to-a-ransomware-attack.

\bibitem{csoonline}
K.~Zurkus, ``Hackers prey on human resources using ransomware,''
  https://www.csoonline.com/article/3112855/hackers-prey-on-human-resources-using-ransomware.html.

\bibitem{formby2017out}
D.~Formby, S.~Durbha, and R.~Beyah, ``Out of control: Ransomware for industrial
  control systems,'' in \emph{RSA conference}, 2017.

\bibitem{nahorney2017internet}
B.~Nahorney, ``Internet security threat report,'' 2017.

\bibitem{proofpoint}
Proofpoint, ``The human factor 2019 report,''
  https://www.proofpoint.com/us/resources/threat-reports/human-factor.

\bibitem{sittig2016socio}
D.~F. Sittig and H.~Singh, ``A socio-technical approach to preventing,
  mitigating, and recovering from ransomware attacks,'' \emph{Applied Clinical
  Informatics}, vol.~7, no.~02, pp. 624--632, 2016.

\bibitem{gallegos2017social}
P.~L. Gallegos-Segovia, J.~F. Bravo-Torres, V.~M. Larios-Rosillo, P.~E.
  Vintimilla-Tapia, I.~F. Yuquilima-Albarado, and J.~D. Jara-Saltos, ``Social
  engineering as an attack vector for ransomware,'' in \emph{Chilean Conf.
  Electrical, Electronics Engineering, Information and Communication
  Technologies}.\hskip 1em plus 0.5em minus 0.4em\relax IEEE, 2017, pp. 1--6.

\bibitem{richardson2017ransomware}
R.~Richardson and M.~M. North, ``Ransomware: Evolution, mitigation and
  prevention,'' \emph{International Management Review}, vol.~13, no.~1, p.~10,
  2017.

\bibitem{marechal2008advances}
S.~Marechal, ``Advances in password cracking,'' \emph{Journal in Computer
  Virology}, vol.~4, no.~1, pp. 73--81, 2008.

\bibitem{nomoreransom}
``No more ransomware project,''
  https://www.nomoreransom.org/en/about-the-project.html.

\end{thebibliography}

\end{document}